\documentclass[aps,pre,preprint,showpacs,showkeys,preprintnumbers,amsmath,amssymb,pr,egroupedaddress]{revtex4}

\usepackage{graphicx}
\usepackage{dcolumn}
\usepackage{bm}

\begin{document}

\title{The effect of tamper layer on the explosion dynamics of atom
clusters}

\author{Zolt\'{a}n Jurek}
\email{jurek-z@szfki.hu}
\author{Gyula Faigel}
\affiliation{Research Institute for Solid State Physics and Optics\\
             H 1525 Budapest, POB 49, Hungary}

\date{\today}

\begin{abstract} The behavior of small samples in very short and
intense hard x-ray pulses is studied by molecular dynamics type
calculations. The main emphasis is put on the effect of various
tamper layers about the sample. This is discussed from the point of
view of structural imaging of single particles, including not only
the distortion of the structure but also the background conditions. A
detailed picture is given about the Coulomb explosion, with
explanation of the tampering mechanism. It is shown that a thin water
layer is efficient in slowing down the distortion of the atomic
structure, but it gives a significant contribution to the background.
\end{abstract}



\maketitle

\section{\label{sec:intro}Introduction}

Today’s most important tool for structure determination is x-ray
crystallography. This method works only on samples having spatial
periodicity. However, there is a growing need for the determination
of the atomic order of small non-periodic entities such as
biomolecules, individual nanoparticles, atomic clusters etc. Since in
these systems the radiation damage is not distributed randomly among
a huge number of identical units (elementary cells in crystals), the
average structure significantly changes with any single damage event.
Therefore the structure determination of these particles requires a
different approach then that of crystalline substances. A solution,
which avoids geometrical distortion was suggested by Solem et al.
\cite{Solem1982,Solem1984} and later it was extended to biomolecules
by Neutze et al. \cite{Neutze2000}.  Their idea was catalyzed by the
construction of future x-ray free-electron lasers (XFELs)
\cite{LCLSstudy,TeslaReport}, which will provide extremely bright
hard x-ray beam with very short pulses. According to their reasoning
elastically scattered photons should be collected during the very
short and intense pulse of XFEL’s. Although in long term the
radiation damage is fatal (the sample explodes), in extremely short
times, the atoms of the sample do not move appreciately. Limiting the
collection of elastically scattered photons to this period we could
obtain useful data for the original structure of the sample. The key
question in this case is the dynamics of the sample explosion: How
much time do we have for imaging? Various models have been worked out
to describe the behavior of the sample in the beam
\cite{JurekEPJD2004,Bergh2004,HauRiege2004}. Although the explosion
dynamics predicted by these models slightly differ in details, but
they all arrive at the same general conclusion: to do successful
imaging with atomic resolution one needs shorter pulses then the
pulse length of the presently constructed sources. Since the
realization of shorter pulses is not simple, there is a search for
methods by which the useful time interval for the measurement of
elastically scattered photons could be extended. A suggestion to
delay atomic motion was put forward by Hau-Riege et al.
\cite{HauRiege2004}. They would surround the particle by a tamper
layer. In the case of biological molecules this layer would naturally
be a water shell.  However, the effect of this layer is not trivial.
Lately, Hau-Riege et al. published model calculations, which examined
this possibility \cite{HauRiege2007}. They concluded that a tamper
water layer delays the explosion of the inner part of the sample, and
they suggested that pulses with width of 50~fs or even larger could
be used for atomic resolution imaging of single bio-molecules. Their
model calculations were based on a 1D continuum approach. Since the
parameter range used in their calculations is not available at the
present technical level, it is very difficult to check the validity
of the various simplifications.  A more detailed and realistic
picture could be given by molecular dynamics modeling. However, the
CPU requirement of these calculations increase very fast with the
number of particles, therefore it is difficult to model large enough
systems. We have developed a special MD algorithm for the description
of the dynamics of a small atom-cluster (or a molecule) in the
intense x-ray pulse \cite{JurekEPJD2004,Faigel2005}.  Lately, we
improved this code to a highly parallelized form. Using this program
we are able to model systems with similar sizes as in Hau-Riege’s
paper. So we did a series of calculations, partly to check the
conclusions of the simpler continuum approach, and partly to get a
more detailed picture of the Coulomb explosion of atom-clusters. 

The paper consist of 6 parts: after the introduction (part
\ref{sec:intro}) we give a brief general description of the explosion
process, describe the main features of our model, and compare our
results with that of Hau-Riege’s (part \ref{sec:compare}). In part
\ref{sec:tlayer} we discuss the effect of the composition of the
tamper layer. In part \ref{sec:pulse} we give an estimate for the
unavoidable background contributions and discuss the effect of pulse
length and shape. In part \ref{sec:classification} we show the result
of model calculations starting from the parameters satisfying the
requirement given by the classification. At last in part
\ref{sec:summary} our results are summarized and conclusions are
drawn. 

\section{\label{sec:compare}Comparison of the explosion dynamics
obtained from the continuum and molecular-dynamics models} 

In the proposed single molecule imaging experiments, individual
identical molecules will be exposed to XFEL pulses one-by-one in
random, unknown orientation. Estimates for the number of elastically
scattered photons show that single diffraction patterns will be very
noisy, due to photon counting statistics. So the compilation of a 3D
diffraction pattern, which is necessary to solve the structure is not
possible directly from the individual 2D images. One has to collect
many pictures into each orientation bins, and add these to improve
statistics. This means that the minimum requirement of photon
counting statistics is determined by the ability to classify the
diffraction patterns according to their orientation. 

Model calculations of the classification process \cite{Bortel2007} give us a
good estimate for the number of photons as a function of the required real
space resolution. Based on the work of G. Bortel and G. Faigel
\cite{Bortel2007} we are considering XFEL pulses with energy of 12~keV and with
fluence in the range of $10^{13}-10^{14}$~photons/pulse. As pulse length we use
10 and 50~fs, which is shorter then the pulse length of XFEL-s under
construction but it is realistically reachable. The reason of using this very
short pulse width is that even under this short time, atoms might move
appreciable distances.  The pulse shape might also influence the imaging of the
atomic structure, therefore we consider flat top and Gaussian shapes. 

After defining the beam parameters we briefly outline the processes governing
the behavior of the sample under the influence of the x-ray pulse.
Approximately $10^{13}$ photons are focused to a 100~nm spot, in which our
sample is situated. Three direct interactions are considered: Compton
scattering, elastic scattering and photoionization. The first interaction is
the smallest, and gives a negligible continuous background. Part of the
elastically scattered photons carry the information on the structure, these are
what we want to select and use later for the structure determination. The other
part contributes to the background (see detailed description later). The
dominant process is however the third, the photoionization. This process leads
to damage. It produces electrons with energy close to the incident photon
energy, and also with lower energy indirectly through the Auger process. The
energy of these slow electrons depends on the elements and for C, N, O which
are the main constituents of biological samples it lies between $250-500$~eV.
The appearance of these electrons is a few fs after excitation.

Both high and low energy electrons scatter elastically and
inelastically on the atoms and ions of the sample. Since the elastic
scattering does not deposit energy in the sample it does not
contribute directly to the damage process. It only modifies the path
of electrons, and through this, it may slightly change the number of
inelastic events. Concerning damage, the inelastic scattering (often
called impact ionization or secondary ionization) of electrons is the
most important. The cross section of this process is large for low
energy electrons. Therefore the contribution of Auger electrons is
dominant in this respect. At later times when electrons slow down
recombination processes also take place. However, these processes
come into play in the developed phase of the Coulomb explosion, so
they are not important from the point of view of imaging. All the
above processes are taken into account in our molecular-dynamics type
model. The detailed description of this model is given in
\cite{JurekEPJD2004}. 

Based on this model we arrived at the following picture of the
Coulomb explosion \cite{JurekEPJD2004}. The main driving force of the
explosion is the K shell photoionization. Photoelectrons leave the
system with $\sim$1/7 light speed leading to a positively charged
particle. Following this the hollow atoms relax through Auger decay,
emitting electrons of a few hundred eV energy. If the system is not
too large (less then hundred Angstrom), most of the Auger electrons
also leave the system in the early time of the pulse. However, as the
positive charge of the particle increases first Auger and later for
large molecules even photoelectrons cannot escape. These non-bound
electrons further ionize atoms by impact ionization resulting even
more free electrons with lower and lower energy. Parallel with these
processes, gradually a rearrangement of the originally homogeneous
charge distribution takes place. An almost neutral inner core and a
highly positively charged outer shell develop. In the inner part, the
free electron cloud screens the Coulomb force between the positive
ions. In an earlier work \cite{JurekEPJD2004} we have shown that the
velocity of the particles in the inner core is much slower then in
the outer shell. As a result the explosion of the core is slower then
the explosion of the shell.  This picture naturally leads to the
application of sacrificial tamper layer, suggested by Hau-Riege et al
\cite{HauRiege2004}. The question is that what type of layer is the
best, and within which conditions.

\begin{figure*} 
\includegraphics[scale=0.45]{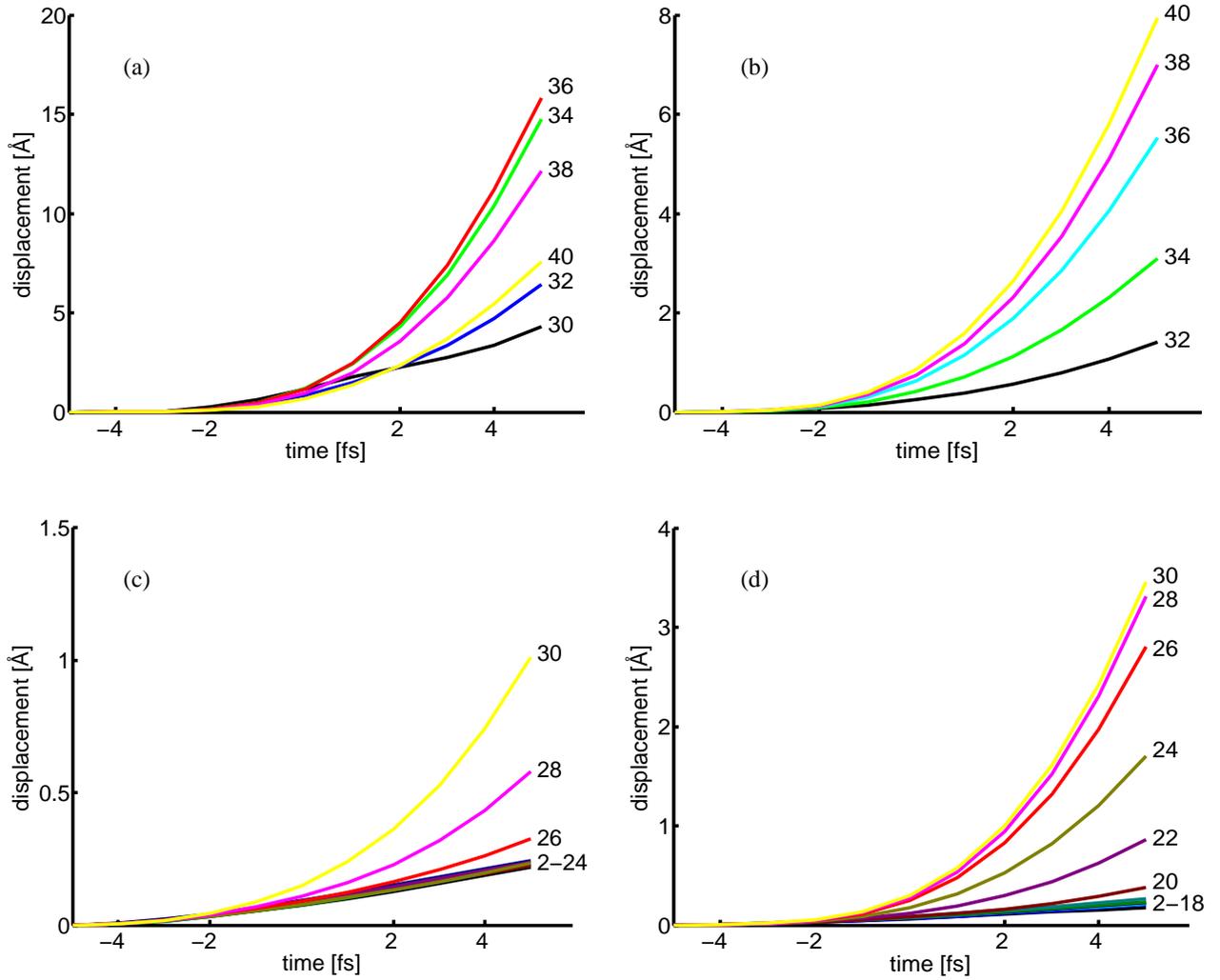}%
\caption{\label{fig:F1}(Color online) Relative displacement of the atomic
shells as a function of time for H (a), O (b), C (c) during the Coulomb
explosion of a small particle composed of a 60~\AA\ diameter sample surrounded
by 10~\AA\ water. The time scale is chosen in such a way that zero time
corresponds to the center of the pulse (half pulse). As a reference the
displacement of C atoms (d) for a 60~\AA\ diameter cluster without the water
tamper layer is also plotted.}
\end{figure*}

First, we compare our results with that of Hau-Riege’s
\cite{HauRiege2007}. In \cite{HauRiege2007} the authors found that
about 10~\AA\ tick water tamper layer is optimum for imaging, since
it effectively slows down the explosion of the inner part and at the
same time it gives the smallest contribution to the elastic x-ray
background. Therefore we performed a calculation for an 80~\AA\
diameter particle, composed of a 30~\AA\ radius core (the sample)
containing C atoms and a 10~\AA\ thick water tamper layer. A 10~fs
flat top x-ray pulse hit the sample. The pulse contained $10^{13}$
12~keV photons focused to a 100~nm spot. In Fig.~\ref{fig:F1} the
relative displacement of the atomic shells are shown as a function of
time for each element (H (a), O (b), C (c)) independently.

As a reference we also plotted on this figure the displacement of C
atoms (d) for a 30~\AA\ radius cluster without the water tamper
layer. It is clear that the explosion of the sample is milder
applying the tamper layer. So our conclusion agrees with that of
Hau-Riege’s. Looking at the numbers, we find that at the end of the
pulse the outer layer of the sample moves about 3~\AA\ in the case of
no tamper layer and 1~\AA\ with tamper layer.  Beside the sample
itself it is interesting to analyze the motion of the tamper layer.
There are characteristic differences between Hydrogen and Oxygen
atoms: first hydrogen moves faster and correspondingly larger
distances, secondly H atomic shells are intermixed with oxygen. The
first observation is trivial since one expects that lighter atoms
move faster, because the q/m ratio (q is the ionization level of the
atom, and m is the mass of the nucleus) is higher for H. However, the
second feature is surprising: H atoms originally located at the
central part of the water shell move faster then atomic shells at
larger distances (Fig.~\ref{fig:F1}.a). This feature can be explained
by the different ionization mechanism of Hydrogen as compared to
Oxygen and Carbon atoms. The photoionization of Hydrogen is
negligible relative to the impact ionization. Therefore slow
electrons generated by the Auger process on Oxygen and Carbon (and
consecutive impact ionizations) ionize more effectively the Hydrogen.
However, the effect of these electrons is the highest in the central
part of the water layer. The reason for this is threefold: i. Oxygen
photoionization cross section is higher then that of Carbon; ii.
Hydrogen atoms are close to Oxygen; and consequently Oxygen atoms
produce more Auger electrons in a shorter time then Carbon. iii. A
large part of Auger electrons produced at the outer edge of the water
shell leave the system without interaction (going radially out). All
these lead to a faster ionization of Hydrogen atoms being in the
central part of the water shell and consequently larger coulomb
forces and faster explosion. 

\begin{figure*} 
\includegraphics[scale=0.4]{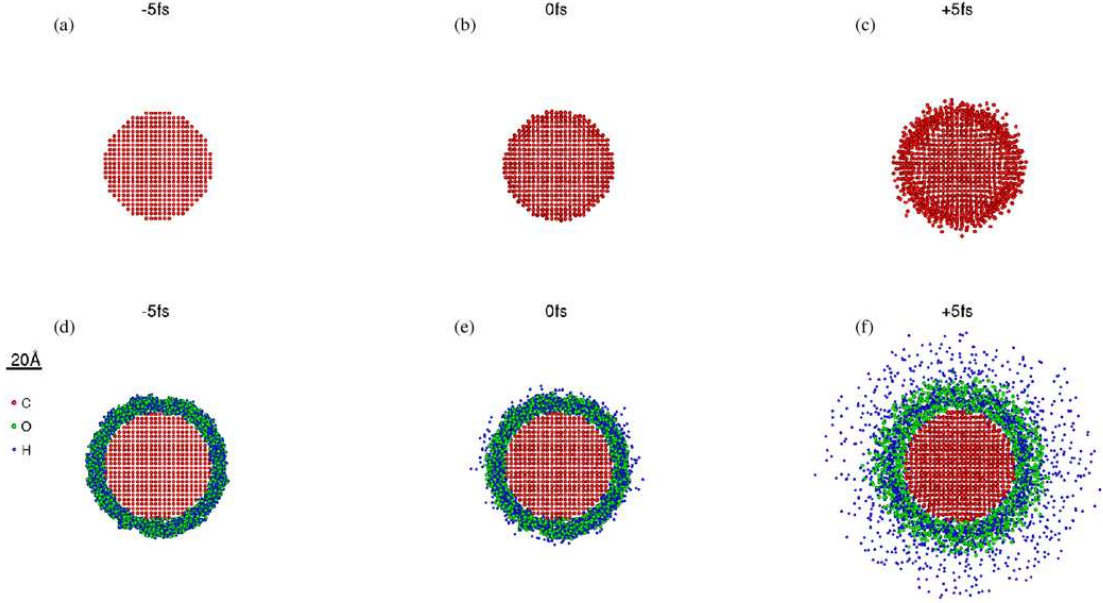}%
\caption{\label{fig:F2}(Color online) Real space image of the atomic
arrangements at different times during the x-ray pulse for the same pulse and
particles as in fig. 1. The atoms and ions are only shown in the cross section
through the center of the sample.  Upper panel: sample without tamper layer at
the beginning of the pulse, at half time and at the end of the pulse (a, b, c),
respectively. Lower panel sample with tamper layer at the beginning of the
pulse, at half time and at the end of the pulse (d, e, f), respectively. } 
\end{figure*}

At the end of this section we show three more characteristics of the
explosion, which naturally comes out from our model, but they are
difficult to derive from the continuum approach. The first is the
real space image of the atomic arrangements at different times
(Fig.~\ref{fig:F2}). We chose three times: the zero time (a) (at the
beginning of the pulse, the original particle surrounded by the water
shell), at half time (b) (when half of the photons in a pulse hit the
sample) and at the end of the pulse (c). 

\begin{figure*} 
\includegraphics[scale=0.5]{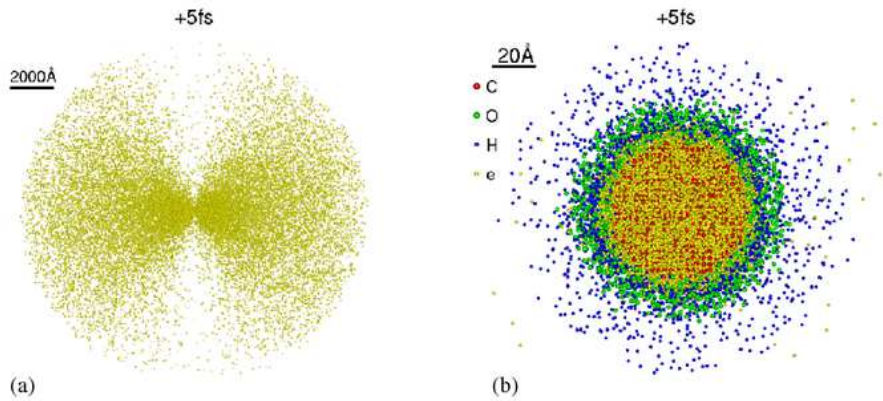}%
\caption{\label{fig:F3}(Color online) The spatial distribution of electrons for
the same particle and pulse parameters as in fig. 1. The sample with the tamper
layer is shown only. The arrangement of electrons is shown at the end of the
pulse in large scale view (a) and in a smaller scale (b). } 
\end{figure*}

We show two sets of images: the original sample without tamper layer
(a, b, c) and the sample with the tamper layer (d, e, f). Looking at
these images one can get a clearer picture of the explosion in real
space. In the upper part of the figure the bare sample is shown. The
development of a thin distorted layer is clearly visible. This points
to the application of sacrificial tamper layer. In the lower part of
Fig.~\ref{fig:F2}, the sample surrounded by a 10~\AA\ thick water
layer is shown. It is clear from the lower right panel that in this
case the original atomic arrangement of the sample does not change
significantly during the pulse. From this figure it is also clear,
that the background coming from the water shell will be a composite
picture; an average of the changing positions of Oxygen and Hydrogen
atoms (ions). The second characteristic is the spatial distribution
of electrons (Fig.~\ref{fig:F3}). 

\begin{figure} 
\includegraphics[scale=0.45]{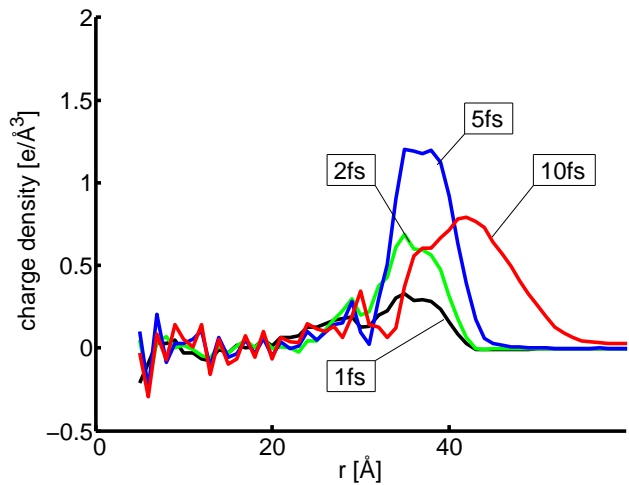}%
\caption{\label{fig:F4}(Color online) Net local charge as a function of the
radial distance from the center of the sample at different times (1, 2, 5, 10
fs). } 
\end{figure}

The 3D arrangement of electrons (yellow spheres) is shown at the end
of the pulse. In Fig.~\ref{fig:F3}.a a large scale view, while on
Fig.~\ref{fig:F3}.b a zoomed image of the electrons are shown. On the
large scale a butterfly shape form composed of photoelectrons and in
the center, at the place of the sample, a small nucleus built up from
Auger and secondary electrons can be seen. The anisotropic
distribution of photoelectrons is a result of the linearly polarized
incident beam. At this scale one cannot distinguish individual
electrons at the center. However, on the zoomed scale in
Fig.~\ref{fig:F3}.b - in contrast to the photoelectron distribution
-, an almost isotropic arrangement of Auger and secondary electrons
is clearly shown. Beside the spatial distribution of electrons, it is
very useful to plot the net local charge as a function of the radial
distance from the center of the sample. This is shown in
Fig.~\ref{fig:F4} at different times during the pulse. 

In a relatively short time from the beginning of the pulse between
$1-2$~fs an almost neutral core surrounded by a positively charged
shell develops. This explains why the explosion of the sample is
delayed by the use of tamper layer. This type of behavior has been
predicted by our earlier calculations for smaller particles
\cite{JurekEPJD2004}.  However, in those calculations there was no
tamper layer, so the outer part of the sample was blown out fast. The
replacement of this outer part by a sacrificial tamper layer is a
natural, good idea.  However, the composition of this layer is not
trivial.

\section{\label{sec:tlayer}The choice of tamper layer}

In this section we examine the effect of the composition of the
tamper layer on the explosion dynamics. The tamper layer basically
acts in two ways: i. it produces electrons, which might diffuse into
the central part (into the sample itself), and ii. the nuclei of the
tamper layer are a barrier for the atoms of the inner part, these
cannot cross this layer easily. 

\begin{figure} 
\includegraphics[scale=0.45]{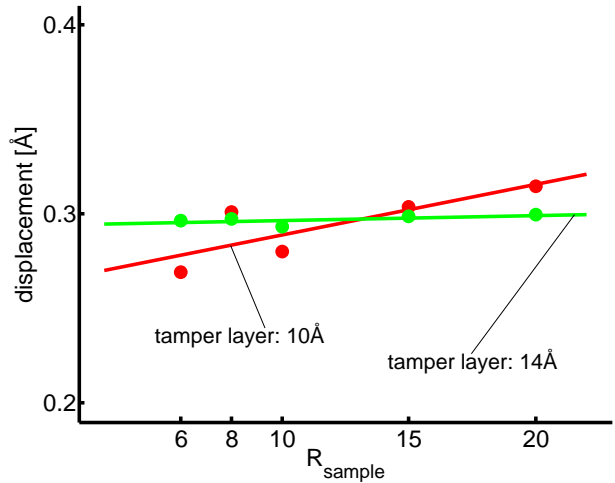}%
\caption{\label{fig:F5}(Color online) Maximum displacements of the outer layer
of the samples as a function of sample size at the end of the pulse for 40~\AA\
diameter samples covered by 10~\AA\ (red curve) and 14~\AA\  (green curve)
water tamper layer. }
\end{figure}

The first effect leads to two things: a.) these excess electrons can
cause secondary ionization of the atoms in the sample. If this
ionization is more effective then the ionization caused by the
electrons produced in the sample part itself the tamper layer might
even speed up the Coulomb explosion, b.) the excess electrons can
also increase the number of free electrons in the central part. These
electrons rearrange so that they help to form the neutral core ie.
they facilitate the screening of the positive charges in the sample.
Depending on the speed of the screening, this effect may slow down
the explosion of the inner part. The balance of these two processes
combined with the barrier effect of the nuclei of the tamper layer
determines the dynamics of the Coulomb explosion. Since both the
barrier effect and the electron production depend on the atomic
number, we carried out model calculations for tamper layers composed
of elements with different atomic number. We chose Helium, Carbon,
Nitrogen, Oxygen, Argon and water. The water does not fit into the
line since it is a two component system. We included it because it is
the most natural cover layer for bio-molecules. Calculations were
done for 20~\AA\ radius samples, covered by layers with thicknesses
chosen so that the total number of electrons in the layers was
approximately the same as we find in a 10~\AA\ thick water layer.
The 20~\AA\ radius was chosen as a compromise between CPU time and
sample size.  However, we emphasize here, that our calculations can
be safely extrapolated for larger samples. To make this statement
more evident, we show the results of a series of calculations for
different sample sizes (6, 8, 10, 15, and 20~\AA\ radiuses) covered
by 10~\AA\ water each.  In Fig.~\ref{fig:F5} (red curve) the maximum
displacements of the outer layer of the samples are shown as a
function of sample size at the end of the pulse. 

\begin{figure} 
\includegraphics[scale=0.45]{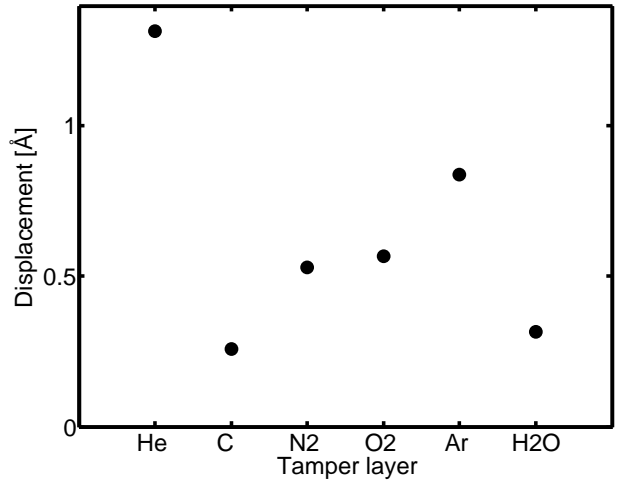}%
\caption{\label{fig:F6} The largest deviation of the sample's outer layer for
various tamper materials. }
\end{figure}

In all cases it remains below 1~\AA\, indicating that the motion of
the atoms themselves does not prevent the solution of the structure
by this precision. However, there is an increase of the displacement
with the size. Extrapolating this to larger distances would lead to
too large atomic displacement for large samples.  However, slightly
increasing the thickness of the water layer (to 14~\AA), and plotting
the size dependence in this case (Fig.~\ref{fig:F5} green curve), we
see that the increase of the displacement becomes much milder,
(almost independent of the sample size). The conclusion is that the
optimum layer thickness of water has to be scaled with the sample
size. However, for smaller samples we should use the smallest
thickness down to 10~\AA\ in order to decrease the elastic x-ray
background, without sacrificing resolution. Turning back to the
original problem of tamper layer composition, we plotted the largest
deviation of the sample’s outer layer for various tamper materials in
Fig.~\ref{fig:F6}. 

There is a minimum at the C, at the sample’s own element. The light
He does not help much because there is negligible photoionization of
He, therefore no excess electrons help the screening of positive
charges of the sample. Further the mass of He atoms is small, they
cannot significantly withhold C atoms from moving. On the other end
of the line at the Argon the photoionization is significant but
together with this there is a very strong secondary ionization of C
atoms caused by the Ar’s Auger and secondary electrons. This speeds
up the explosion of the sample outer layers, and this is not
compensated by the larger mass of Ar atoms, because the atomic
density of Ar is very low, so they cannot withhold the Carbon atoms.
The effect of the water layer is close to that of the pure carbon.
The reason is that the Auger electrons of O are very effective in the
secondary ionization of Hydrogen, therefore many excess electrons are
traveling into the Carbon sample. Even though these ionize the Carbon
atoms by secondary ionization, the mass of the Oxygen atoms combined
with the large atomic density of oxygen compensate for this and slow
down effectively the explosion.

\section{\label{sec:pulse}The effect of pulse parameters and tamper
layer on the background}

\begin{figure*} 
\includegraphics[scale=0.41]{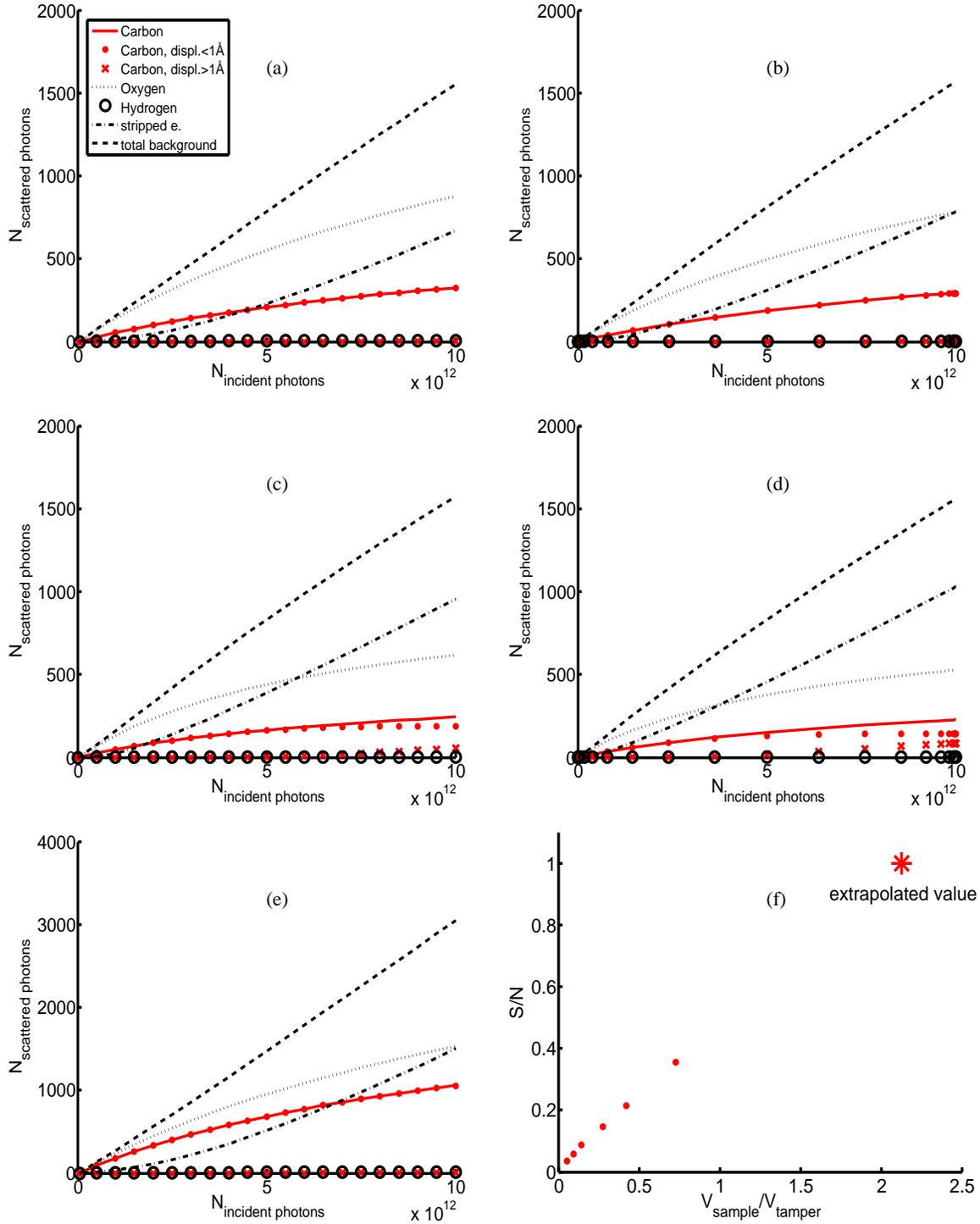}%
\caption{\label{fig:F7}(Color online) Background contributions and the useful
signal of a 20~\AA\ radius carbon sample surrounded by a 10~\AA\  thick water
layer as a function of the integrated number of incident photons for flat top
(a, c) and Gaussian (b, d) shape pulses with 10 (a, b) and 50 (c, d) fs widths.
Part (e) shows the same contributions as the previous parts for a 30~\AA\
radius sample covered by 10~\AA\  water in a flat top 10 fs wide pulse. In Part
(f) the signal to noise ratio as a function of the sample volume normalized
with the tamper layer volume is depicted. An extrapolated value is marked by a
star. } 
\end{figure*}

In previous sections we analyzed the x-ray pulse induced motion of
the atoms. However, this is not the only problem of imaging. 2D
diffraction patterns taken during consecutive pulses have to contain
enough information for successful classification, ie. for finding
patterns of particles arriving into the beam with the same
orientation. This step is crucial, since without it the assembly of
3D diffraction pattern fails, and consequently the structure solution
is not possible. The success of classification is strongly background
dependent. Therefore in this section we give an estimate of the
background of an ideal measurement. The sources of unavoidable
background are: elastically scattered photons by the free electrons,
elastic scattering from the atoms ions of the tamper layer, elastic
scattering of those atoms ions of the sample which moved more then
the tolerance (the required resolution). The first two factors starts
with the pulse and is integrated over the pulse duration following
the time evolution of the number of free electrons and the changing
scattering power of the ions of the tamper layer. The third factor is
different; it gives some contribution only close to the end of the
pulse. It adds a relatively small value to the background but at the
same time it decreases the signal with the same amount. This
contribution depends on the duration and shape of the pulse. We
calculate these contributions for 10 and 50~fs flat top and Gaussian
shape pulses. As a model system we take a 20~\AA\ radius carbon
sample surrounded by a 10~\AA\ thick water layer. In
Fig.~\ref{fig:F7} (a, b, c, d) we plot all three contributions as a
function of the integrated number of incident photons for flat top
and Gaussian shape pulses with 10 and 50~fs widths. 

For comparison we also plotted the useful signal, the elastic
scattering from the sample. It is clear that the largest part of the
background is coming from the tamper layer. The second largest is the
free electron contribution and the smallest part is coming from the
deteriorating part of the sample. The total background is about three
times larger then the signal. We can get a more promising picture by
taking larger samples maintaining the same thickness for the tamper
layer. In this case, the relative contribution of the tamper layer
decreases. For illustration we show the background contributions for
a 30~\AA\ sample with 10~\AA\ water (Fig.~\ref{fig:F7}.e). It is very
instructive to plot the signal to noise ratio as a function of the
sample volume normalized with the tamper layer volume
(Fig.~\ref{fig:F7}.f). For 10~fs flat top pulses we obtain a linear
dependence, and can safely extrapolate to larger sample sizes. We get
S/N~1 for $\sim$70~\AA\ radius samples. Further increase of the
sample size results in an improved signal to noise ratio. However,
for longer pulses (Fig.~\ref{fig:F7} c, d) -especially for the
Gaussian shape-, the linear extrapolation does not hold. The reason
for this is that in this case the background from the deteriorating
part of the sample gets appreciable and at the same time the useful
signal decreases significantly, lowering the signal to noise ratio.
For a 50 fs Gaussian pulse the signal to noise ratio drops to about
0.06 for a 40~\AA\ diameter sample covered by 10~\AA\ water. This
might cause significant problems in the classification and
reconstruction process. 

\section{\label{sec:classification}Modeling based on classification
requirements}

\begin{figure*} 
\includegraphics[scale=0.5]{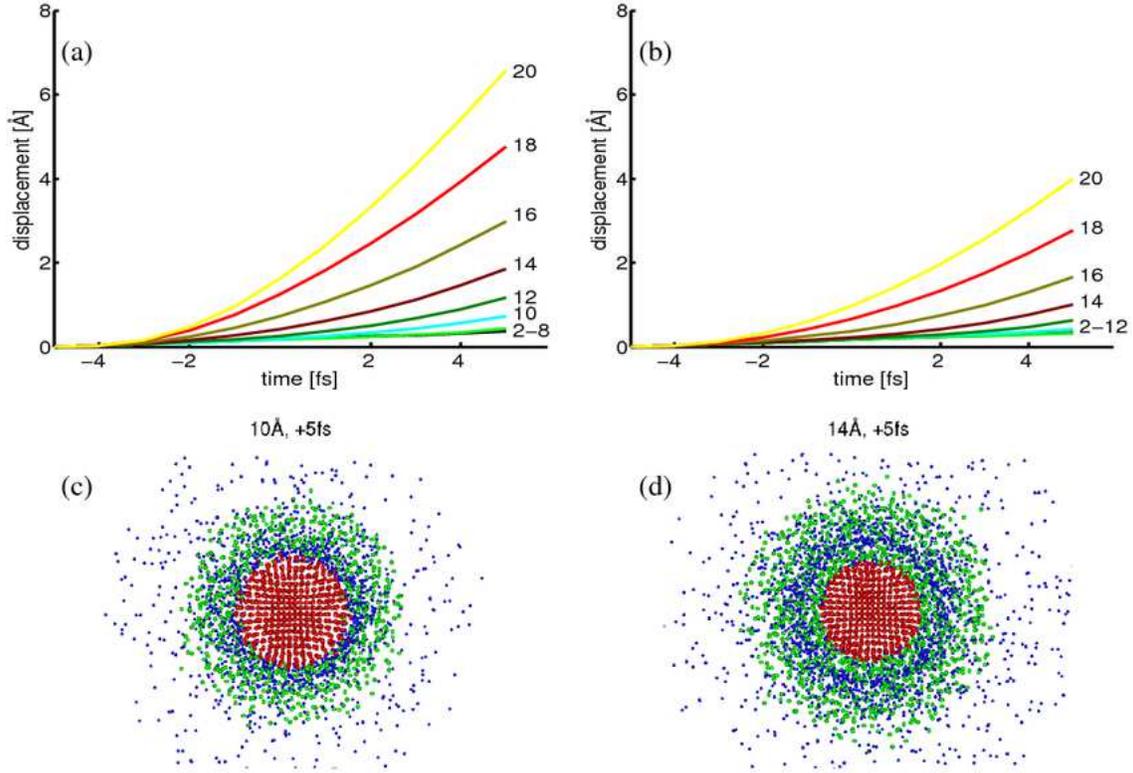}%
\caption{\label{fig:F8}(Color online) The time evolution of atomic shells of a
40~\AA\  diameter sample surrounded by a 10~\AA\  (a) and 14~\AA\  (b) water
shell for high fluence. The corresponding real space images at the end of the
pulse are shown on parts (c) and (d), respectively.  } 
\end{figure*}

So far we concentrated on the dynamics of explosion without analyzing
whether the number of useful elastically scattered photons were
enough to obtain any meaningful information on the structure. In this
section we take into account the requirements given by the first step
of data analyses the classification. Numerical modeling shows, that
these requirements are the strictest among the steps of structure
solution so the classification is the bottle neck in the evaluation
process. Lately, G. Bortel and G. Faigel estimated the minimum number
of photons necessary for classifications at various sample sizes
\cite{Bortel2007}. Based on their results we did a series of
calculations with more realistic input parameters. Classification
considerations indicated the need of higher fluence on the sample.
This can be reached either by using tighter focusing or by higher
input intensity. We choose the second option and used $10^{14}$
photons in a pulse with flat top shape leaving the focal spot 100 nm.
The sample had 40~\AA\ diameter surrounded by a 10~\AA\ water shell.
The time evolution of atomic shells is shown on Fig.~\ref{fig:F8}.a. 

The deterioration of the sample is drastic. The displacement of the
outer shell is about 8~\AA\ at the end of the pulse. So imaging with
atomic resolution is not possible. However, in section
\ref{sec:tlayer} we have seen that a small increase of the thickness
of the tamper layer decreases the explosion of the core. Therefore we
repeated the calculation for 14~\AA\ water tamper layer. The
displacement of the atomic shells is shown in Fig.~\ref{fig:F8}.b.
There is significant improvement; the largest deviation decreased two
half compared to the 10~\AA\ tick tamper layer. However even in this
case we will have a degraded resolution. For illustration of the
distortion we can expect we show the real space image of the atomic
structure for the two tamper layers in Fig.~\ref{fig:F8}.c and d.
Further, we should also keep in mind that the price we pay for the
reduced distortion is a tripled background caused by the larger
amount of water.

\section{\label{sec:summary}Summary and conclusion}

We have examined the effect of tamper layers on the dynamics of
Coulomb explosion of small Carbon particles in the XFEL beam. We
found that a 10-14~\AA\ water layer decrease the deterioration of the
sample by a factor of three. The optimum thickness depends on the
sample size and incident intensity. One should use thicker layer for
larger size and for higher intensity. We also did a series of
calculations for sacrificial layers composed of various elements (He,
C, N, O, Ar). We found that very light (like He) and heavy (like Ar)
elements are not effective in slowing down the Coulomb explosion of
the sample. However, carbon, nitrogen and oxygen help in retaining
the original atomic arrangement of the samples. The best result is
shown by carbon. The effect of water layer is similar to the pure
carbon case. Beside the explosion dynamics we also studied the
various background contributions to the diffraction pattern of the
sample. This is crucial from the point of the view of imaging the
atomic structure. We found that the largest contribution is coming
from the tamper layer. However, the weight of this contribution is
decreasing for larger samples. We found that in the case of 10 fs
flat top pulses the background is smaller then the signal, for
samples larger then 70~\AA\ in radius. However, for longer pulses the
situation gets less favorable. For example the signal to noise ratio
decreases to $\sim$0.06 for a 40~\AA\ diameter sample illuminated by
a 50~fs Gaussian pulse. This might be crucial concerning the
possibility of single molecule imaging, since the first step of
structure solution -the classification-, depends on the achievable
statistics (SN ratio) reachable by a single shot. 

Based on the requirements of the classification process we also
modeled the Coulomb explosion at higher photon/pulse value (at
$10^{14}$ ph/pulse) then the design parameters of XFEL-s under
construction. We found that the maximum displacement of atoms
increases to about 8~\AA\ (with 10~\AA\ water layer) but it can be
pushed down to 4~\AA\ by increasing the thickness of the tamper
layer. This may allow close to atomic resolution imaging. However, we
pay for this with a significantly increased background i.e. a
decreased SN ratio.  Therefore further detailed numerical studies on
the classification has to be done and probably new less noise
sensitive classification methods has to be worked out. Our
calculations indicate that in future experiments one has to do
detailed preliminary modeling of the full imaging process in order to
find the optimum experimental conditions for a given sample.

\begin{acknowledgments}The work reported here was supported by OTKA
67866 and NKFP1/0007/2005 grants. We thank Gabor Bortel and
Mikl\'{o}s Tegze for the illuminating discussions.
\end{acknowledgments}





%
%

\end{document}